\documentclass{aastex61}

\usepackage{graphicx,times}             
\usepackage{natbib}                     
\usepackage{amssymb,amsmath}            
\bibpunct{(}{)}{;}{a}{}{,}

\newcommand\aastex{AAS\TeX}

\usepackage{ulem}

\received{\today}
\revised{XXX XX, 2020}
\accepted{XXX XX, 2020}

\submitjournal{ApJ}

\shorttitle{\aastex\ Pulsar Candidate Sifting Using MICNN}
\shortauthors{H.-T. Lin, \& X.-R. Li \& Q.-G. Zeng}

\begin{document}

\title{Pulsar Candidate Sifting Using Multi-input Convolution Neural Networks}

\correspondingauthor{Xiangru Li}
\email{xiangru.li@gmail.com}

\author[0000-0002-0786-7307]{Haitao Lin}
\affil{School of Mathematical Sciences,
South China Normal University,
Guangzhou, 510631, China}
\affil{School of Mathematics and Statistics,
Hanshan Normal University,
Chaozhou, 521041, China}

\author{Xiangru Li}
\affiliation{School of Computer Science,
South China Normal University,
Guangzhou, 510631, China}
\author{Qingguo Zeng}
\affiliation{School of Mathematical Sciences,
South China Normal University,
Guangzhou, 510631, China}



\begin{abstract}

Pulsar candidate sifting is an essential process for discovering new pulsars. It aims to search for the most promising pulsar candidates from an all-sky survey, such as High Time Resolution Universe (HTRU), Green Bank Northern Celestial Cap (GBNCC), Five-hundred-meter Aperture Spherical radio Telescope (FAST), etc. Recently, machine learning (ML) is a hot topic in pulsar candidate sifting investigations.
However, one typical challenge in ML for pulsar candidate sifting comes from the learning difficulty arising from the highly class-imbalance between the observation numbers of pulsars and non-pulsars.
Therefore, this work proposes a novel framework for candidate sifting, named multi-input convolutional neural networks (MICNN). The MICNN is an architecture of deep learning with four diagnostic plots of a pulsar candidate as its inputs.
To train our MICNN in a highly class-imbalanced dataset, a novel image augment technique, as well as a three-stage training strategy, is proposed.
Experiments on observations from HTRU and GBNCC show the effectiveness and robustness of these proposed techniques.
In the experiments on HTRU, our MICNN model achieves a recall of 0.962 and a precision rate of 0.967 even in a highly class-imbalanced test dataset.

\end{abstract}




\section{Introduction} \label{sect:intro}
The study on pulsars has promoted the development of fundamental physics and astronomy, especially in the field of gravitational waves detection \citep{taylor1994binary} and dark matter discovery \citep{baghram2011prospects}.
Many pulsars were discovered in some modern pulsar projects such as HTRU \citep{keith2010high,levin2013high}, the pulsar Arecibo L-band feed array \citep[PALFA; ][]{deneva2009arecibo}, the low-frequency array tied-array all-sky survey \citep[LOTAAS; ][]{coenen2014lofar}, etc. According to the Australia Telescope National Facility \citep[ATNF; ][]{manchester2005australia}, about 2800 pulsars have been identified so far. However, more new pulsars are expected to be identified and therefore, modern surveys (projects) have been conducted such as the Square Kilometre Array \citep[SKA; ][]{smits2009pulsar} and Five-hundred-meter Aperture Spherical radio Telescope \citep[FAST; ][]{nan2006five,nan2011five,nan2016fast}.
With the advent of astronomical instruments, these surveys will produce a vast amount of data of pulsar candidates.
However, most of them are Radio Frequency Interference (RFI) signals or other non-pulsar noises. Only a fairly small proportion (one in ten thousand at most; \citet{lyon2013study}) of these candidates are from real pulsars which the researchers concern about. Therefore, a significant and necessary task is to separate the real pulsar signals from the non-pulsar ones in a survey, which is known as \textit{candidate sifting} or \textit{candidate classification}.

Recently, pulsar candidate sifting schemes based on machine learning (ML) attract the attention of researchers for their accurate performance and effective computational ability on massive data.
\citet{eatough2010selection} first built a model of artificial neural networks (ANN) for pulsar candidate classification with 12  features. Then, \citet{bates2012high} constructed another ANN model with 22 features. To improve the performance of the ML model, \citet{morello2014spinn} used only 6 empirical features to design a model called Straightforward Pulsar Identification using Neural Networks (SPINN), while \citet{lyon2016fifty} proposed another 6 new statistical features to construct a real-time learning classifier named Gaussian Hellinger Very Fast Decision Tree \citep[GHVFDT;][]{lyon2014hellinger}.
Recently, convolutional neural network (CNN) has also been applied to pulsar candidate sifting. CNN was first used by \citet{zhu2014searching} to construct the Pulsar Image-based Classification System (PICS) whose inputs are four diagnostic images of the pulsar candidates. To further develop PICS, \citet{wang2019pulsar_a} designed a mode PICS-ResNet by replacing the CNNs with Residual Networks (ResNets).
 Then, \citet{guo2019pulsar} proposed a framework involved in a deep convolution generative adversarial network (DCGAN) and achieved satisfactory results. Machine learning methods, especially schemes based on CNNs, have shown significant application in the pulsar candidate sifting.

Pulsar candidates from the surveys are class-imbalanced. In fact, the number of real pulsar candidates is much lower than that of the non-pulsars. This imbalance has negative effects on the performance of machine learning sifting methods. Therefore, methods of data augmentation or oversampling are designed to balance the categories. Besides, the architecture of a model and its training strategy can significantly affect the performance of the candidate sifting. Considering the above-mentioned factors, we propose a novel framework as well as a data augmentation technique and its training strategy for candidate sifting.
Firstly, a CNN-based model, called \textit{Multi-Input Convolutional Neural Networks} (MICNN), is described. Given the chief diagnostic evidence for pulsar identification, the inputs to the MICNN are the sub-integration plot (or the sub-band plot), the folded profile and the DM curve (Fig. \ref{Fig:pulsar-fourfeature}).
Then, to balance the positive samples and negative samples in the training set, a data augmentation technique is designed, namely, \textit{Transforming Image and Adding Gaussian Noise} (TIAGN). Finally, to train the MICNN, a three-stage strategy of training was carried out with the imbalance ratio (IR) increasing.
The evaluation on both HTRU and GBNCC shows that the proposed scheme is capable of pulsar candidate sifting even if the candidates are highly class-imbalanced.

Related works and the used data sets are introduced in Section \ref{sec:data_relatedworks}. A novel generative algorithm TIAGN is given to deal with the class imbalance problem in Section \ref{sec:data_augmentation}. In Section \ref{sec:Methods}, our model of MICNN is presented, while Section \ref{sec:experiments} describes the details of the experimental process. In Section \ref{sec:results}, the experimental results are given and analyzed. Conclusions and discussions of this work are made in Section \ref{sec:discussion_conclusion}.

\section{The data sets and related works}\label{sec:data_relatedworks}
This section introduces two data sets used in this paper and the related works on pulsar candidate sifting based on machine learning.
\subsection{Pulsar candidates}
\label{sect:Dataset}
We will evaluate our approach on both HTRU and GBNCC. HTRU \citep{keith2010high} is an all-sky survey for pulsars and radio transients. The Southern Hemisphere is being observed with the Parkes Radio Telescope, while the Northern Hemisphere is being observed with the Effelsberg 7-beam system. The HTRU Medlat Data set is a collection of labeled pulsar candidates from the intermediate galactic latitude part of the HTRU survey, which was released to the public domain \citep{morello2014spinn} and have been
widely used in pulsar candidate sifting investigations \citep{bates2012high, morello2014spinn, lyon2016fifty, guo2019pulsar, wang2019pulsar_a, wang2019pulsar_b, xiao2020pulsar, lin2020pulsars}. The data set consists of 1196 pulsar signals and 89996 non-pulsar signals.
GBNCC pulsar survey is an ongoing search for pulsars and dispersed pulses of radio emission, using the Robert C. Byrd Green Bank Telescope in West Virginia \citep{stovall2014green}. The survey data set consists of 90008 pulsar candidates. Among them only 56 are real pulsars and 221 are harmonics, which are labeled as pulsar signals in our experiments. Thus, candidates from both HTRU and GBNCC are highly class-imbalanced.

\begin{figure*}
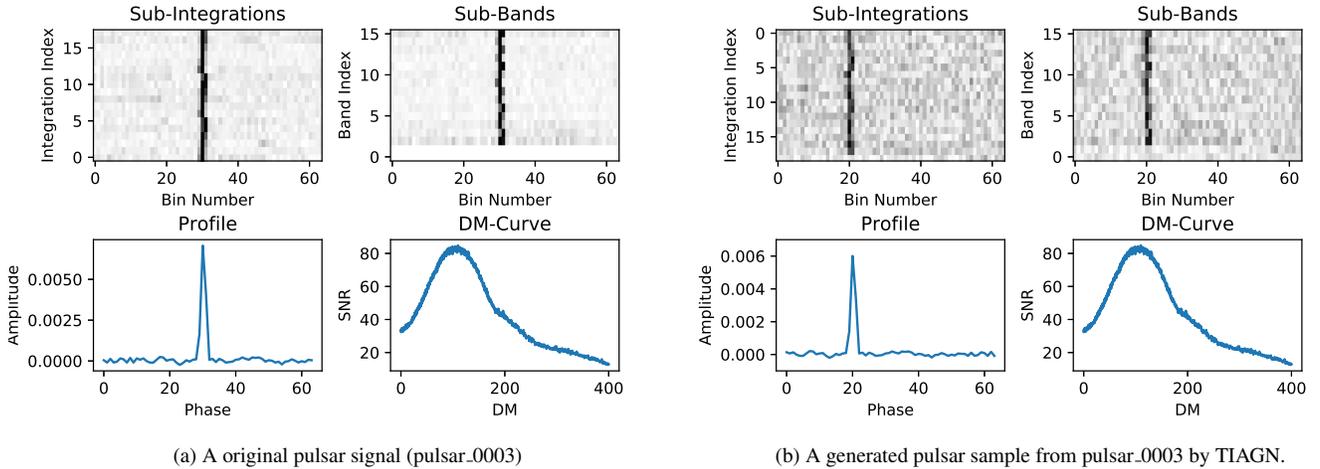

\gridline{\fig{pulsar_fourfeature.eps}{0.5\textwidth}{(a) A original pulsar signal (pulsar\_0003)}
          \fig{pulsar_transform.eps}{0.5\textwidth}{(b) A generated pulsar sample from pulsar\_0003 by TIAGN.}
          }
\caption{Four diagnostic plots of a pulsar candidate on HTRU (left) and its generated signal by TIAGN (right).  Compared with the original pulsar signal in (a), the generated pulsar candidate has the same folded profile in shape (only a shift in peak) and shares the same DM curve, but has more noise in both the sub-integration plot and sub-band plot.}\label{Fig:pulsar-fourfeature}
\end{figure*}

Pulsar candidates are computed by a complex process including RFI removal \citep{keith2010high}, dispersion measure (DM) searching \citep{lorimer2009radio} and period searching.
Fig. \ref{Fig:pulsar-fourfeature} (a) shows four diagnostic plots for a candidate from HTRU.
 A sub-integration plot (on the top left) is obtained by adding signals in all frequency and stacking it by its derived period, while a sub-band plot (on the top right) is generated by summing the signals over all periods in each sub-band of frequency. A folded profile of the candidate (on the bottom right) can be obtained by folding the sub-integration plot or sub-band plot. Ideally, for a pulsar signal, we should see a peak in the folded profile, where the power is much stronger than the noise level. And a DM curve (on the bottom left) is the trial result of the integrated column density of free electrons between the receiver and the signal. For a pulsar signal, the DM-curve should peak at a non-zero value which is regarded as an approximated dispersion measure of the candidate. Since these four plots are the main diagnostic pieces of evidence for pulsar detection, many researchers extracted handcrafted features from them as the inputs into a classifier. Some of the artificial features are easy to be computed and successful for separating the pulsars from the non-pulsars. For instance, \citet{lyon2016fifty} extracted eight statistical features from the folded profile and the DM curve of a candidate. Among them, Pf$_{\mu}$, Pf$_{\sigma}$, Pf$_{k}$ and Pf$_{s}$
 are the mean, the standard deviation, the kurtosis and the skewness of the folded profile, while DM$_{\mu}$, DM$_{\sigma}$, DM$_{k}$ and DM$_{s}$ are the corresponding measures from the DM curve of a candidate. Since these features were extracted easily and have shown to be helpful to separate noise and non-noise candidates in \citet{lyon2016fifty}, we will use them as part of the inputs into our model.

\subsection{Machine learning on pulsar candidate sifting}
This work investigated the pulsar candidate sifting problem based on some supervised machine learning (SML) methods \citep{mitchell1997machine}, which are associated with a set of labelled data.
The idea of SML is to learn a function mapping the features (inputs) of the instances to their corresponding labels.
As for candidate sifting, we aim to train a model mapping the features of the candidates to their categories: pulsars (positives) or non-pulsars (negatives).

According to the existing literatures on pulsar candidate sifting, there are two main types of SML approaches. One type is the models with handcrafted features \citep{eatough2010selection, bates2012high, morello2014spinn, lyon2016fifty, lyon2014hellinger, tan2018ensemble, xiao2020pulsar, lin2020pulsars}. In these models, features are extracted from their diagnostic plots and statistical information of the candidates. For example, \citet{morello2014spinn} designed 6 features for their SML model such as SNR of the folded profile, the ratio between barycentric period and dispersion measure, and the persistence of a signal through the time domain. The other type of SML method takes the diagnostic images of candidates as its inputs. \citet{zhu2014searching} firstly used these diagnostic images as inputs into PICS. They designed PICS with a two-layers architecture.
In the first layer, ANN, CNN and Support Vector Machine \citep[SVM;][]{cortes1995support} were used to rate how "pulsar-like" each of the diagnostic plots was, while in the second layer, a logistic regression (LR) was used with the outputs from the first layer as inputs.  Then, \citet{wang2019pulsar_a} improved the PICS by replacing CNNs with residual networks.
Recently, \citet{guo2019pulsar} used DCGAN to balance the pulsars and non-pulsars in the training set and combined it with SVM and enhance the accuracy of the automatic pulsar candidate identification. \citet{zeng2020concat} designed a Concat Convolutional Neural Network (CCNN) to identify the candidates collected from the FAST data using the four diagnosic plots as inputs and improved candidate sifting performance evidently. Therefore, SML methods have been successfully applied in pulsar candidate sifting.

\subsection{The class-imbalanced problem}
One typical characteristic of the survey data is the highly class-imbalance, as the ratio of non-pulsars to pulsars is 75:1 on HTRU and even 324:1 on GBNCC.
Classification algorithms probably suffer from the problem of training difficulty by being skewed towards one class.
This is known as imbalanced data classification. Imbalanced data classification poses a challenge for predictive modeling.
The class-imbalanced data can decrease the pulsar candidate sifting performance.
Generally, classification algorithms are sensitive to an imbalanced data set and lead to performance losses \citep{longadge2013class,ali2015classification}. Experimental results of \citet{lyon2013study} indicate that less and less true pulsar signals can be correctly judged as the class imbalance increases.
The class-imbalanced data tend to result in a model that predicts each candidate as a non-pulsar and therefore misses the real pulsar signals.

To alleviate this bias and improve the performance of a classification model, one way is to balance the training data by some sampling techniques such as oversampling or undersampling. Take pulsar candidates for example,
the simple oversampling method balances the dataset by duplicating the pulsar signals randomly and keeping the non-pulsar candidates, while the undersampling method balances it by selecting a small number of instances from the non-pulsars signals randomly and keeping the pulsars signals.
Both simple oversampling and undersampling can create balanced dataset for training. However, a model based on simple oversampling probably result in an over-fitting model which is too consistent with the data of the training set and predicts future observations with relatively low reliability. And undersampling method will discard too much useful or important information of non-pulsar signals and raise the false positive rate of a model. Therefore, other methods of dealing with this imbalanced problem were proposed. For example, \citet{wang2019pulsar_b} developed an artificial synthesis method of oversampling by a linear combination of three randomly selected pulsar signals. Based on these generated candidates, they train their CNN model using a class-balanced data set and test the learned model using a class-imbalanced set. \citet{guo2019pulsar} produces deep features of the pulsar signals by the DCGAN technique to balance their training set and thus raise the performance of their models.

\section{Data Augmentation}\label{sec:data_augmentation}

\begin{figure*}[ht!]
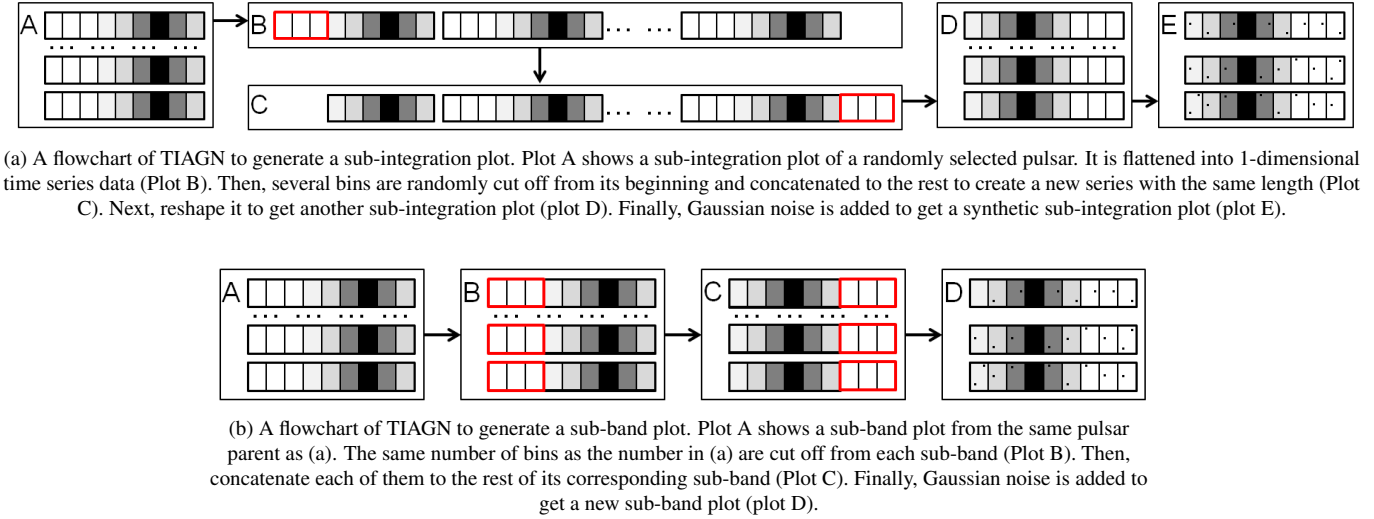

\gridline{\fig{legend.eps}{1\textwidth}{(a) A flowchart of TIAGN to generate a sub-integration plot. Plot A shows a sub-integration plot of a randomly selected pulsar. It is flattened into 1-dimensional time series data (Plot B). Then, several bins are randomly cut off from its beginning and concatenated to the rest to create a new series with the same length (Plot C). Next, reshape it to get another sub-integration plot (plot D). Finally, Gaussian noise is added to get a synthetic sub-integration plot (plot E).}
          }
\gridline{\fig{legend2.eps}{0.7\textwidth}{(b) A flowchart of TIAGN to generate a sub-band plot. Plot A shows a sub-band plot from the same pulsar parent as (a). The same number of bins as the number in (a) are cut off from each sub-band (Plot B). Then, concatenate each of them to the rest of its corresponding sub-band (Plot C). Finally, Gaussian noise is added to get a new sub-band plot (plot D).}
          }
\caption{Flowcharts of TIAGN. (a) The computation process of a sub-integration plot, (b) The computation process of a sub-band plot.}\label{Fig:TIANG}
\end{figure*}

To balance the pulsar candidates without loss of any information of the data, a novel data augmentation approach is proposed and referred to as Transforming Image and Adding Gaussian Noise (TIAGN).  Each artificial candidate from TIAGN consists of four diagnostic figures: a sub-integration plot, a sub-band plot, folded profile plots and DM curve. Each generated sample is transformed from a randomly chosen pulsar. Fig. \ref{Fig:TIANG} shows the flowcharts of TIAGN. Above all, the process of a synthetic sub-integration plot in Fig. \ref{Fig:TIANG} (a) is displayed as follows: First, a randomly chosen plot of sub-integrations (Plot-A) is flattened into 1-dimensional time series data (Plot-B); Next, several bins were cut off randomly from the beginning of these time series data, and concatenated to the end of the rest to create a new series (Plot-C) with the same length; And then, reshape it to get another sub-integration plot (Plot-D); Finally, Gaussian noise is added to get a new sub-integration plot (Plot-E). As for sub-band plot, bins of the same number are cut off from each sub-band and concatenated to the end of its corresponding sub-band. Finally the Gaussian noise is added in a similar way (Fig. \ref{Fig:TIANG} (b)).
The folded profile of this synthetic signal can be obtained by averaging its sub-integrations, while its DM curve keeps unchanged in shape since it originates in the same pulsar.
Fig. \ref{Fig:pulsar-fourfeature} (b) shows a synthetic candidate from pulsar\_0003 (Fig. \ref{Fig:pulsar-fourfeature} (a)) by the TIAGN. Compared with the original pulsar signal in Fig. \ref{Fig:pulsar-fourfeature}, the generated one almost has the same folded profile in shape (only a shift in peak) and shares the same DM curve, but has more noise in both sub-integration plot and sub-band plot.

Different from other artificial synthesis methods, TIAGN algorithm generates ``pulsar-like" signals in a natural way. Compared with the original parents, they just vary in their initial phase due to the translation operation, and should be lower SNR because of the Gaussian noise.
Theoretically, signals produced by TIAGN have kept almost all of the physical and statistical characteristics of their original parents, and therefore these artificial signals can be also considered as some positive data for training. To show the effectiveness of TIAGN, some comparative experiments are illustrated in Section \ref{sect:conclusion:data_aug}.

\section{Model}\label{sec:Methods}

This section describes the proposed MICNN which is a model based on CNNs.
CNNs are feedforward neural networks with convolution operations and have been widely used and proved to be successful in pattern recognition systems, recommendation systems, and other applications. A model based on CNNs often involves some basic types of layers such as convolutional layers, downsampling layers (pooling layers), and fully connected layers (dense layers). Some researchers have constructed models based on CNNs to sift the pulsar candidates and most of them achieved satisfactory performances in their specific surveys \citep{zhu2014searching, guo2019pulsar, wang2019pulsar_a, wang2019pulsar_b, zeng2020concat}.

\begin{figure}[ht!]
   \centering
   \includegraphics[width=10cm, angle=0]{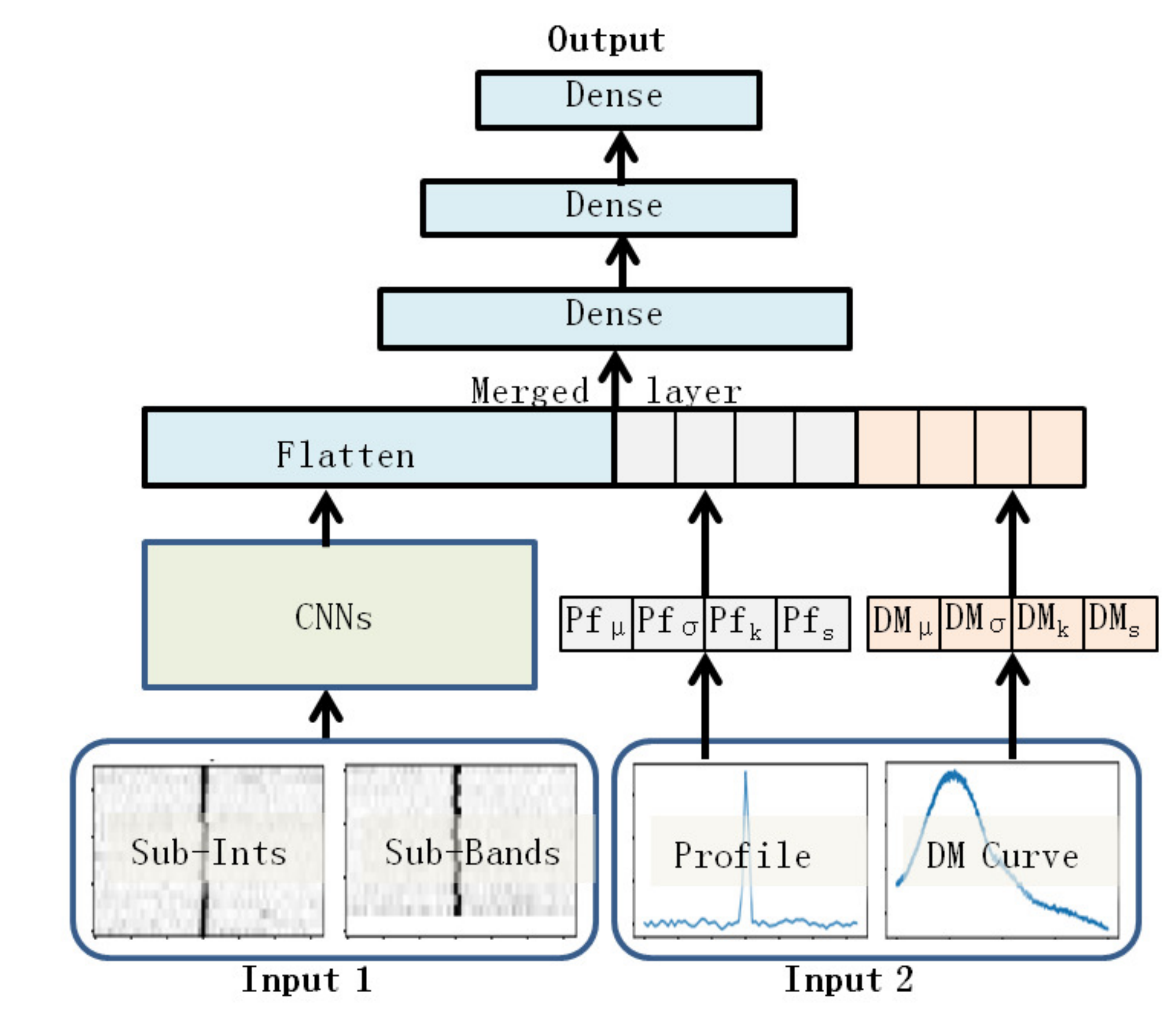}
   \caption{The architecture of MICNN in our work.
   The main input (Input 1) is the sub-integration plot (sub-band plot) and the auxiliary input (Input 2) consists of eight statistical features extracted from the folded profile and DM curves. A merged layer connects the deep features from Input 1 with the eight statistical features from Input 2. See Table \ref{Tab:hyperparameters} for the architecture of MICNN in detail.}
   \label{ModelStructure}
   \end{figure}

This work designs a novel model for pulsar candidate sifting based on CNNs with multiple inputs (Fig. \ref{ModelStructure}). We call it \textbf{Multi-Input Convolution Neural Networks} (MICNN).
There are two inputs in MICNN, a main input and an auxiliary input.
The main input (denoted as \textbf{Input 1}) into MICNN is the sub-integration plot or the sub-band plot of a pulsar candidate. Input 1 is fed into CNN-based networks with a structure of three ``conv-conv-pool"-like blocks. The auxiliary input (denoted as \textbf{Input 2}) into MICNN is a vector extracted from the folded profile and the DM curve, which consists of eight statistical features designed by \citet{lyon2016fifty}.
Then, the output of the CNN-based networks is flattened and concatenated with Input 2. We call this concatenating layer as \textbf{the merged layer} (see Fig. \ref{ModelStructure} ).
Thus, this merged layer consists of two feature sources for a candidate. One comes from the deep features of the sub-integrations or the sub-bands. The other one represents the features from the folded profile and DM curve.
Therefore, the inputs to MICNN take more information compared with some other models whose input is only the sub-integration plot or the sub-band plot.
Finally, this merged layer is followed by three dense layers. In this work, MICNN-1 denotes a MICNN model with the sub-integration plot as its Input 1, while MICNN-2 represents a MICNN model with the sub-band plot as its Input 1.

\begin{table}[ht!]
        \begin{center}
        \caption {Architecture of the MICNN for HTRU data. It takes two inputs and consists of 13 layers, including 6 convolutional layers, 3 pool layers, 1 merged layers, and 3 dense connected layers.} \label{Tab:hyperparameters}

        \begin{tabular}{lllrllllc}
        \hline\noalign{\smallskip}
         No.&Layer name  &Kernel &Size  & Activation   \\
         \hline
            &\textbf{Input 1}     & & $16 \times 64$  &           \\
           1&Convolution &$3\times 3$     & $\textbf{32}\times 16\times 64$  &  PReLU           \\
           2&Convolution &$3\times 3$     & $\textbf{32}\times 16\times 64$  &  PReLU           \\
           3&Max Pooling     &$2\times 2$     & $\textbf{32}\times 8\times 32$  &             \\
           4&Convolution &$3\times 3$     & $\textbf{64}\times 8\times 32$  &   PReLU          \\
           5&Convolution &$3\times 3$     & $\textbf{64}\times 8\times 32$  &   PReLU          \\
           6&Max Pooling     &$2\times 2$     & $\textbf{64}\times 4 \times 16$  &             \\
           7&Convolution &$3\times 3$     & $\textbf{128}\times 4 \times 16$  &   PReLU          \\
           8&Convolution &$3\times 3$     & $\textbf{128}\times 4 \times 16$  &   PReLU          \\
           9&Max Pooling     &$2\times 2$     & $\textbf{128}\times 2 \times 8$  &             \\
           10&Flatten+\textbf{Input2}     &                    & $2048+8$  &             \\
           11&Dense &    & \textbf{512}     & PReLU            \\
           12&Dense        &    & \textbf{128}     & PReLU            \\
           13& Dense       &   & $\textbf{2}$  & Softmax            \\
             & \textbf{Output}       &   &   $2$          &            \\
        \noalign{\smallskip}\hline
        \end{tabular}
        \end{center}
\end{table}

Table \ref{Tab:hyperparameters} shows the hyperparameters in MICNN which are obtained by grid searching method and 10-fold cross validation.
We unify the size of Input 1 to be $16\times 64$ on HTRU, where $16$ is the sub-integration (sub-band) index, while $64$ is the bin index.
The size of sub-integrations (sub-bands) on HTRU is almost $16\times 64$ and thus we keep most of the original information of the data without any transformation. However, the size of Input 1 is changed to $64\times 64$ on GBNCC because of the prescribed size of the sub-integrations (sub-bands) on GBNCC.
There were 6 convolutional layers in MICNN with the increasing channel numbers of 32,32,64,64,128,128, the identical kernel size of $3\times 3$ and, the same activation function of PReLU. One max pooling layer is added between two nearest convolution layers and the stride of Max pooling is 1. Besides, to avoid overfitting, the dropout layers are added to each convolutional layer as well as each dense layer. The dropout rates are 25\% and 50\%, respectively.

\section{Experimental process}\label{sec:experiments}
\label{sect:Results}
This section describes the experimental details. In particular, a training strategy is designed to train MICNN on a highly-imbalanced data set.
\subsection{Performance Measure}
To evaluate the performance of a model, several measures should be computed including the recall, the precision and $F_1$ score \citep{lin2020pulsars}.
The $recall$ rate, the $precision$ rate and $F_1$ score are the primary metrics to measure the quality of a pulsar candidate sifting model. In pattern recognition tasks with binary classification, $recall$ measures how many pulsars could be correctly identified as pulsars from all the real pulsars and, $precision$ measures how many true pulsars would be predicted correctly out of the candidates identified as pulsars. Often, there is an inverse relationship between $recall$ and $precision$, where it is possible to increase one at the cost of reducing the other. Therefore, $F_1$ score, defined as the harmonic mean of $recall$ and $precision$, is a trade-off between them.

\subsection{Data Preprocessing}
We split the dataset into three parts: training set (55\%), validation set (20\%) and test set (25\%). The training set is used to fit a model, while the validation set provides an evaluation of a model being trained to tune its hyperparameters. The test set is used only to assess the performance of the final model.
To evaluate the performance of trained models with different levels of class imbalance, the training set was resampled by our proposed oversampling approach TIAGN. Here, the level of class imbalance is measured as the ratio of non-pulsars to pulsars, which is called \textit{imbalance ratio} (IR). We denote \textit{Training set 0} as the original training data, and denote \textit{Training set 1}, \textit{Training set 2} as two training sets generated by TIAGN with ratio 5:1 and 20:1, respectively (Table \ref{Tab:sample}). To verify the generalization ability of our model, we further evaluated the MICNN on GBNCC, a highly class-imbalanced data set with IR 324:1. The IRs of the three-stage training are 6:1, 40:1 and 324:1 respectively in the GBNCC application.

\begin{table}[ht!]
        \begin{center}
        \caption {The configuration of the data sets based on HTRU. \textit{Training set 0} is the original training dataset, while \textit{Training set 1}, \textit{Training set 2} are generated from \textit{Training set 0} by TIAGN. IR: imbalance ratio.}\label{Tab:sample}
        \begin{tabular}{lllllll}
        \hline\noalign{\smallskip}
         Part & Pulsar & Non-pulsar & Total & IR   \\
         \noalign{\smallskip}
        \hline\noalign{\smallskip}
          Training set $0$    & 655   & 49499 & 50154 &75:1\\
          Training set $1$   &9900    & 49499 & 59399 &5:1\\
          Training set $2$   &2475    & 49499 & 51974 &20:1\\

          Validation set& 240    & 18000 & 18239 &75:1\\
         Test set    & 301    & 22497 & 22798 &75:1\\
         \noalign{\smallskip}
        \hline
        \end{tabular}
        \end{center}
\end{table}

\subsection{Training Strategy}\label{Sect:TrainingStrategy}

Different from models with single input, the MICNN is contrained using two loss functions. One is the main loss function based on sub-integration (sub-band). The other is the auxiliary loss from the statistical features.
To compile the MICNN, we chose \textbf{Adam} as the optimizer and \textit{binary cross entropy} as the loss function.
Adam is an algorithm based on first-order gradient-based optimization.

However, like many classification algorithms, the MICNN suffers from a class imbalance problem. It failed to train in \textit{Training set $0$} (Section \ref{sect:conclusion:onestage}). The reason is that MICNN is accuracy-driven and therefore a model trained using a highly class-imbalanced training set tends to result in a high accuracy but a low recall rate. However, the latter metric is what we more concern about. To deal with this problem, a three-stage training strategy is designed for the MICNN.

\begin{figure}
   \centering
   \includegraphics[width=\textwidth, angle=0]{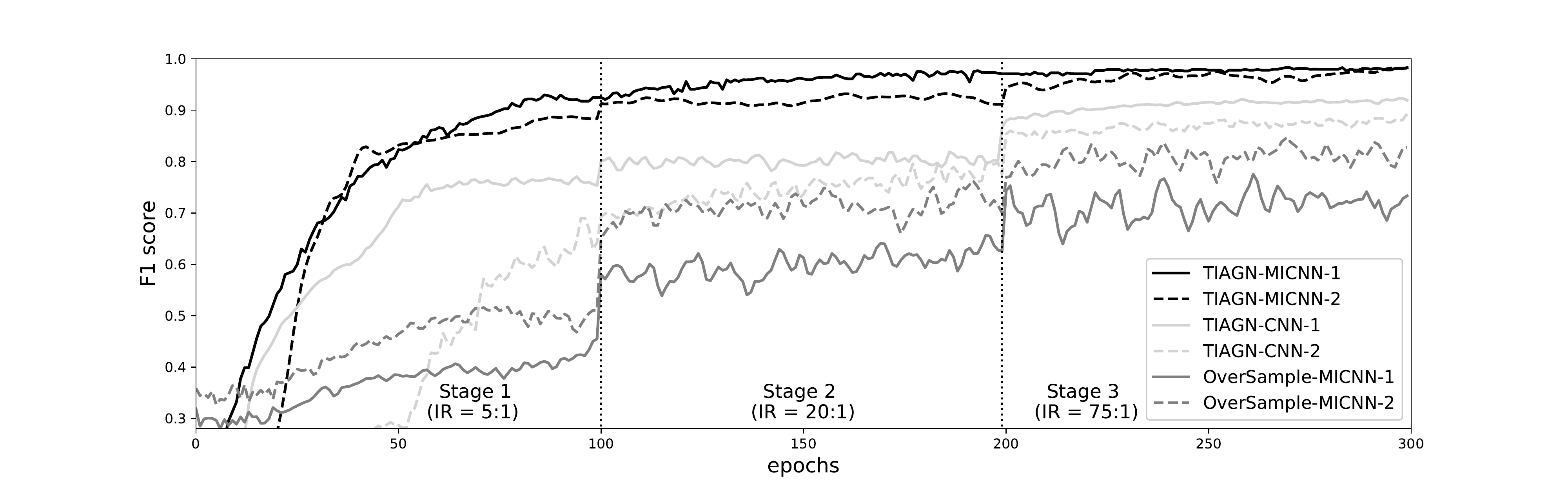}
   \caption{The F1 score of six models on the same validation dataset in the three-stage training process.
     We trained MICNN-1 (MICNN-2) and CNN-1 (CNN-2) on the TIAGN data to see the additional effects of the auxiliary input (Input 2). Here, CNN-1 (CNN-2) is the CNNs architecture on the MICNN-1 (MICNN-2) with a single input (Input 1).
      Besides, to show the effectiveness of the TIAGN method, we also train MICNN-1 (MICNN-2) on the simple over-sampling data. All the models are trained with three-stage training strategy. Suffix `1' means the model trained on the sub-integration plots, and `2' means the model learned on the sub-band plots.}\label{Fig:training_process}
   \end{figure}

In the first stage, we train the MICNN using \textit{Training set $1$} which is not so imbalanced and result in a model $M_1$ when the loss of validation set is stable. Then, we load $M_1$ as an initial model in the second stage where we train the MICNN using \textit{Training set $2$}. It results in another model $M_2$. Likewise, we use $M_2$ as an initial model in the third stage where we train the MICNN using \textit{Training set $0$}. We gradually increase the imbalance ratios of the training data in the learning process, from 5:1 to 20:1 and finally to 75:1 on the HTRU data. Once the model converges in the former training set, it is used to initialize the next training stage. In this way, the MICNN can be successfully trained using \textit{Training set $0$} and result in a model $M_0$, achieving better $F_1$ score than the other two models. The training process of MICNN were recorded in Fig. \ref{Fig:training_process} and the performance of these models mentioned are presented in Table \ref{Tab:threestage}.

In our training process, the former model is used to initialize the following one. The idea of our training strategy for the MICNN is a technique referred to transfer learning \citep{west2007spring}. Transfer learning focuses on storing knowledge gained in solving one problem and applying it to another which is related. Initializing a network with transferred features can improve the performance \citep{yosinski2014transferable}.

\subsection{Experimental settings}
Two kinds of MICNN with different inputs were implemented on HTRU. Among them, MICNN-1 denotes the MICNN whose Input 1 is the sub-integration plot, while MICNN-2 denotes the MICNN whose Input 1 is the sub-band plot. Each MICNN follows the training strategy (see Section \ref{Sect:TrainingStrategy}) with three levels of class imbalance (
Table \ref{Tab:sample}). Similar experiments were conducted on GBNCC, as we altered the imbalance rates of the three-stage training sets to 6:1, 40:1, and 324:1 respectively.

To investigate the effect of class imbalance on the performance of the MICNN, the recall, the precision and the $F_1$ score are calculated (Table \ref{Tab:threestage}). Besides, to evaluate the role which the eight extra features play on, CNN-1 and CNN-2 are tested in the same process, where CNN-1 and CNN-2 are the CNN architectures on the MICNN-1 and the MICNN-2 with a single input (Input 1), correspondingly (Fig. \ref{Fig:training_process}). All the models were tested using the same test set (Table \ref{Tab:sample}). Besides, to show the effectiveness of the TIAGN method, we also compared the performances of MICNN based on TIAGN with those based on the simple over-sampling data. More related discussions are made in Section \ref{sect:conclusion:data_aug}.

\section{Results}\label{sec:results}

\subsection{Performance and analysis}
\begin{figure}
  \centering
  \includegraphics[width=9cm]{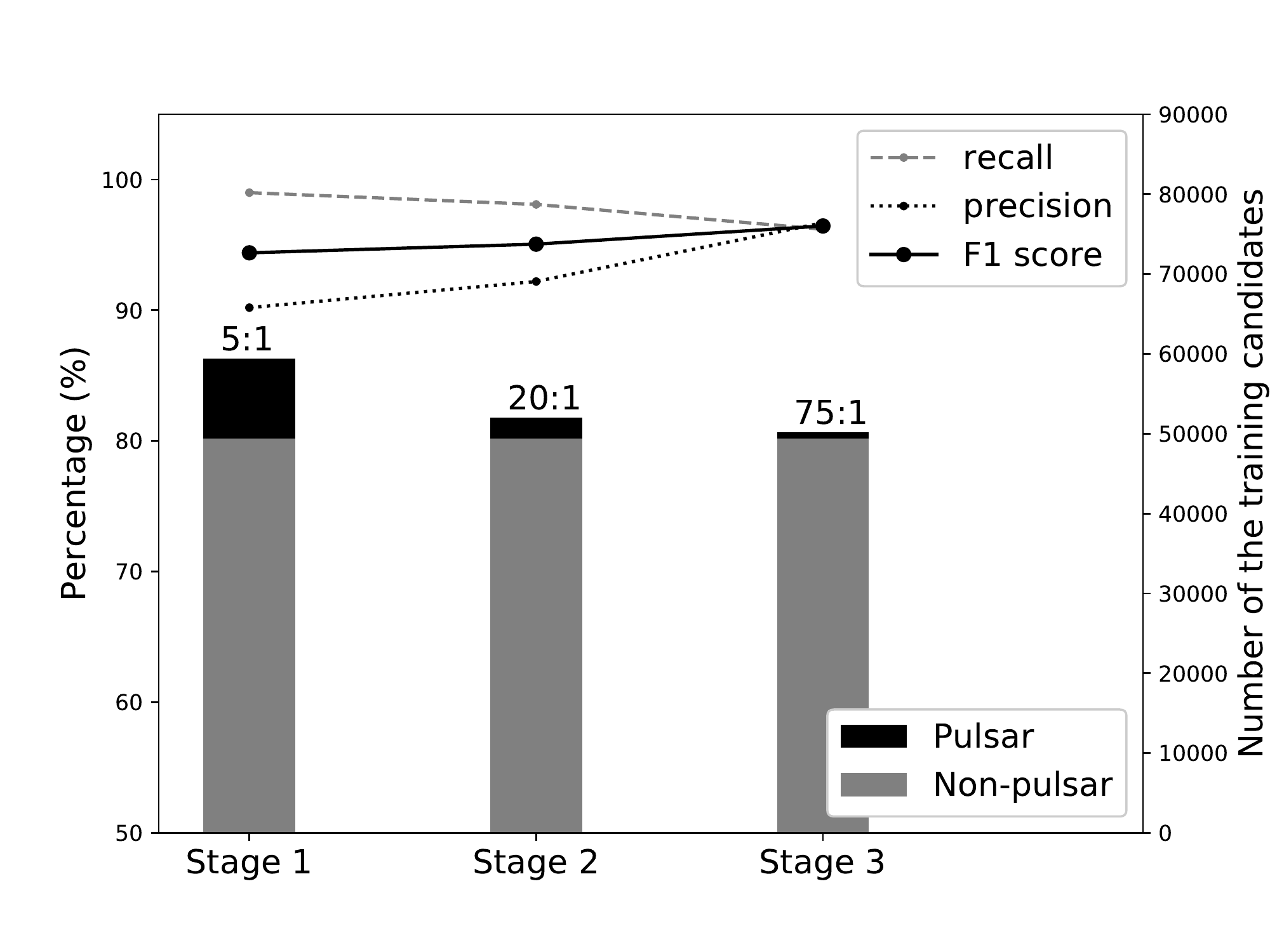}\\
  \caption{The performance of MICNN-1 on HTRU with the three-stage training strategy. The recall rate, precision rate and the $F_1$ score are evaluated on the same validation set, while the balance ratio varies from 5:1 to 20:1 and finally to 75:1. The results show that the recall is a bit decreasing while the precision and the $F_1$ score increases as the imbalance ratios getting large.  }\label{Fig:performance_subint}
\end{figure}

Following the above-mentioned training strategy, the performances of our models on HTRU and GBNCC were given in Table \ref{Tab:threestage}.
The recall rate, precision rate and the $F_1$ score of the MICNN are evaluated on the same validation set. As the imbalance ratios increase, the recall rate is a bit decreasing while both the precision and the $F_1$ score are increasing. These can be observed in Fig. \ref{Fig:performance_subint} where the performance metrics of MICNN-1 were evaluated on HTRU. The model achieved less misjudged pulsars but more false positives when MICNN was trained using \textit{Training set $1$}. However, given the performance of $F_1$ score, models trained using \textit{Training set $0$} were better than the other two cases. The reason is that the test set is similar with the \textit{Training set $0$} on IR and imbalance characteristics.
Also, we noticed that for HTRU survey, MICNN-1 performs a bit better than MICNN-2, while for GBNCC survey, MICNN-2 is better in each case of different class imbalance ratios when $F_1$ score is the chief measure of the model. It implies that features extracted from the sub-integration plots are more relevant than those from the sub-band plots for HTRU, while the sub-band plots are more trustworthy for GBNCC.

 \begin{table}[ht!]
        \begin{center}
        \caption{Performance of the MICNN in each stage with the three-stage training strategy on HTRU survey and GBNCC survey, respectively. We train MICNN-1 and MICNN-2 with the IR increasing. Here MICNN-1 denotes the MICNN whose Input 1 is the sub-integration plot, while MICNN-2 denotes the MICNN whose Input 1 is the sub-band plot. All the models were tested using the same test set in Table \ref{Tab:sample}.}\label{Tab:threestage}
        \begin{tabular}{lllllllll}
        \hline\noalign{\smallskip}
         Survey&Training set & IR & Method & Recall & Precision & $F_1$ Score   \\
         \noalign{\smallskip}
        \hline\noalign{\smallskip}
        HTRU&Training set $1$ &5:1  &MICNN-1     &\textbf{0.990}  &0.902  & 0.943\\
                               &&&MICNN-2   &\textbf{0.970}  &0.875   &0.920\\
        &Training set $2$ &20:1 &MICNN-1     &0.980          &0.922  &0.951\\
                                &&&MICNN-2   &0.965          &0.930  &0.947\\
        &Training set $0$ &75:1 &MICNN-1     &0.962  &\textbf{0.967}  &\textbf{0.965}\\
                                &&&MICNN-2   &0.956  &\textbf{0.968}  &\textbf{0.962}\\
         \noalign{\smallskip}
        \hline
         GBNCC&Training set $1$ &5:1  &MICNN-1    &\textbf{0.894}   &0.347  &0.503\\
                                &&&MICNN-2  &\textbf{0.939}   & 0.712 & 0.810\\
        &Training set $2$ &40:1 &MICNN-1    &0.833   & 0.578 & 0.684\\
                                &&&MICNN-2  &0.909   &0.822   &0.863\\
        &Training set $0$ &324:1 &MICNN-1    &0.773  &\textbf{0.729}  &\textbf{ 0.750}\\
                                &&&MICNN-2    &0.884  &\textbf{0.913}  &\textbf{ 0.898}\\
         \noalign{\smallskip}
        \hline
        \end{tabular}
        \end{center}
\end{table}

Furthermore, we compared the performance of MICNNs with their single-input CNNs (Table \ref{Tab:results:unbalance}) on HTRU. That's to say, to verify the effectiveness of multiple inputs in MICNN, the performance of their according CNNs models whose input is only the sub-integration plot or the sub-band plot. The results show that MICNN performs better than CNNs. The mean validation $F_1$ score is 0.965 for the MICNN-1 and 0.920 for the CNN-1, while 0.962 for the MICNN-2 and 0.869 for the CNN-2. These results indicate that the eight extra statistical features from the folded profile and the DM curve are sufficiently beneficial to improve the performance of MICNN.

Meanwhile, performances are compared between our MICNNs and other models on HTRU.
 The results show that our MICNN models achieved a high recall rate of $\sim 0.96$, as well as a high precision of $\sim 0.97$ (see Table \ref{Tab:results:unbalance}). One reason for the improvement of the performance is that there are more pieces of information for the input to our MICNN model. The MICNN considers sub-integrations (the sub-bands) as well as the other eight statistical features, while the DCGAN-L2-SVM \citep{guo2019pulsar} and DCNN-S \citep{wang2019pulsar_b} only took the sub-integration plot or the sub-band plot as their input. Another reason may be the fact that MICNNs were trained in a class-imbalanced set whose IR is the same as the test set, while others were trained using a class-balanced set and tested using a class-imbalanced set. In fact, a model can predict better if the distribution of the training data is more similar to that of the test data.

Finally, to verify the effectiveness of TIAGN, comparative experiments on HTRU were carried on between MICNNs trained with TIAGN candidates and MICNNs trained with simple over-sampling. The results in Table \ref{Tab:results:unbalance} showed that TIAGN dose significantly improved the performance measure of MICNNs. We explained the reasons in Section \ref{sect:conclusion:data_aug}.

 \begin{table}[ht!]
        \begin{center}
        \caption {Performance of different methods on HTRU.
        MICNN-1 denotes the MICNN whose Input 1 is the sub-integration plot, while MICNN-2 denotes the MICNN whose Input 1 is the sub-band plot. CNN-1 and CNN-2 are the CNN architectures on MICNN-1 and MICNN-2 with a single input (Input 1). To show the effectiveness of the TIAGN method, MICNN-1 (MICNN-2) trained on the simple over-sampling data were evaluated in the experiments of this table.}\label{Tab:results:unbalance}
        \begin{tabular}{lllllllll}
        \hline\hline\noalign{\smallskip}
         Reference &Balance Technique& Model & Recall & Precision & $F_1$ Score   \\
         \noalign{\smallskip}
        \hline\noalign{\smallskip}
          Wang et al.(2019b)  &Linear Combination& DCNN-S         & 0.962 & 0.963 &0.962\\
          Guo et al.(2019)&DCGAN&DCGAN-L2-SVM-1  &\textbf{0.966}& 0.961&0.963\\
                          &DCGAN&DCGAN-L2-SVM-2  &\textbf{0.963}& 0.965 &\textbf{0.964}\\

         Our work  &{TIAGN}&CNN-1   &	0.892&	0.950&	0.920\\
                   &{TIAGN}&CNN-2  &  0.842&	0.898 &	0.869\\
                   &{TIAGN}&MICNN-1&	0.962&	\textbf{0.967}&	\textbf{0.965}\\
                          &{TIAGN}&MICNN-2& 0.956& \textbf{0.968} & 0.962\\
                   &{OverSample}&{MICNN-1}&{0.764}   &{0.786} &	{0.775}\\
                   &{OverSample}&{MICNN-2}&{0.810}   &{0.883} & {0.845}\\
         \noalign{\smallskip}
        \hline
        \end{tabular}
        \end{center}
\end{table}

\subsection{Misclassified Pulsars}
\begin{figure*}
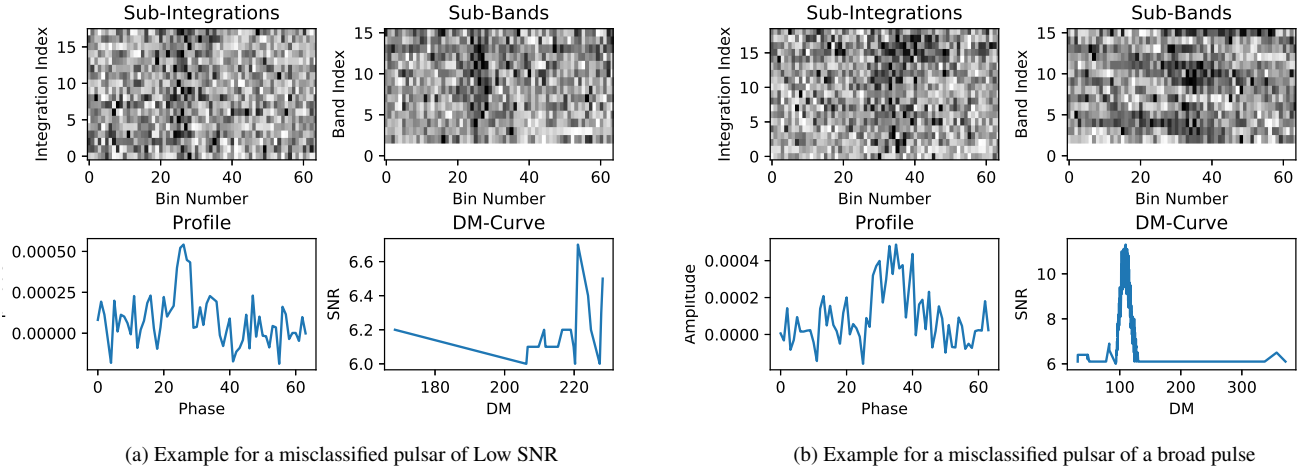

\gridline{\fig{Mis_pulsar_0029.eps}{0.5\textwidth}{(a) Example for a misclassified pulsar of Low SNR}
          \fig{Mis_pulsar_1175.eps}{0.5\textwidth}{(b) Example for a misclassified pulsar of a broad pulse}}
\caption{Diagnostic plots of two instances of misclassified pulsars: (a) Pulsar\_0029 on HTRU, (b) Pulsar\_1175 on HTRU.}\label{Fig:Mis_pulsar}
\end{figure*}

In this section, the misjudged pulsar candidates are concerned about.
 Although MICNN have achieved a high the recall rate on both HTRU and GBNCC, some real pulsar signals were misclassified, about 3 on HTRU and 9 on GBNCC. According to their plots (Fig. \ref{Fig:Mis_pulsar}), descriptions of these misclassified pulsars can be made as follows.

 \begin{itemize}
\item[i)]\textit{Low SNR.} It can be inferred from both their sub-integration plots and sub-band plots that the SNRs of the misclassified pulsar candidates were very low.
    The noises of these candidates in the off-pulse region are strong, while their signals in the on-pulse region are too weak to be recognized (see Fig. \ref{Fig:Mis_pulsar} (a)).
 \item[ii)]\textit{Broad pulse.} The folded profile of this type of misjudged pulsars is heavily scattered in the off-pulse region {and} shows a broad pulse in its folded profile (see Fig. \ref{Fig:Mis_pulsar} (b)).
\end{itemize}


\section{Discussions and Conclusions}\label{sec:discussion_conclusion}
\label{sect:conclusion}

In this study, we propose a deep learning model named MICNN for pulsar candidate sifting. It is a CNN-based architecture with multiple diagnostic plots of a pulsar candidate as its inputs.
Different from the existing CNN models for pulsar detection, the MICNN was trained following a three-stage training strategy with imbalance ratios of the training sets gradually increasing. To balance the training sets, a natural data augmentation technique of oversampling called TIAGN was proposed to create more positive signals in the first and second stages of the training process.
Experiments on HTRU and GBNCC show that MICNN, as well as its training strategy, can be helpful for pulsar prediction despite the highly class-imbalanced training data.

\subsection{The merged layer in MICNN}
One of the contributions of our work is to create a merged layer in MICNN to connect deep features with handcrafted features. Different from other CNN-based models, MICNN combines two kinds of features in a model by a merged layer. One includes deep features from a CNN-based model whose input is the sub-integration plot (the sub-band plot) of a candidate. The other one consists of eight statistical features extracted from the folded profile and the DM curve.
In fact, some researchers have also used multiple features as the inputs to their proposed models. For example, \citet{zhu2014searching} and \citet{wang2019pulsar_a} proposed PICS and PICS-ResNet for candidate sifting, respectively. However, both the PICS and PICS-ResNet system are two-step structurally. In the first step, they train four models. Each model with an independent diagnostic plot as input. And then a discriminant model was used in the second step by combining the outputs of the first step as its inputs. However, our framework is quite different from the PICS or the PICS-ResNet, as the merged layer enables our MICNN architecture to join together the multiple features from the diagnostic plots and makes the training process of MICNN to be an end-to-end process instead of a two-step one.

\subsection{Data augmentation of TIAGN}\label{sect:conclusion:data_aug}
To address the imbalance problem, TIAGN is proposed as a data augmentation technique, which consists of two procedures on a randomly chosen pulsar observation: an image translation and some Gaussian noise disturbances. The TIAGN not only helps avoid overfitting but also augment data in a way to improve the performance of our classifier.

The additional information carried about by the TIAGN is expressed in the generated candidates with different positions of peaks. These candidates together with their original parents enable our MICNNs to learn the invariant features from sub-integrations (sub-bands) and thus improve the performance measures of the models. In fact, image translation in TIAGN is of great significance for training a CNN model. Several authors have shown that: small translations of the input image can drastically change the network’s prediction \citep{azulay2018deep}, as they explained that the reason lies in the subsampling operations in the CNN architecture. As for the candidates from TIAGN, some candidates with different positions of peaks may actually come from the same original parent. This information (knowledge) can be easily discovered by MICNN when the training set is enriched by TIAGN. Therefore the model of MICNN tends to learn the invariant features of a pulsar from both these generated candidates and their parents. In the final step of TIAGN, some Gaussian noises were added to the candidates. The intension of noise adding is to improve the generalization ability. It is a common technique for generalization in a deep learning model to add some extra noise to the input of a model.

To evaluate the effect of TIAGN, comparative experiments (Table \ref{Tab:results:unbalance}) are carried out between the TIAGN and the simple oversampling which is a simple method to address the class-imbalanced problem by making multiple copies of pulsar candidates. We have trained two kinds of MICNN models with the three-stage training strategy. The only difference between them is their training set, as one was from the TIAGN method and the other was from random oversampling. Fig. \ref{Fig:training_process} recorded the training process of these models. It shows that models based on TIAGN are more stable and accurate, achieving $\sim$0.96 of $F_1$ score, while the models based on sample oversampling suffer from the problem of abnormal fluctuations and got lower $F_1$ score, which indicates the models of the later do not generalize well on their validation data.

\subsection{Advantages of the three-stage training strategy}\label{sect:conclusion:onestage}
\begin{figure}
  \centering
  \includegraphics[width=18cm]{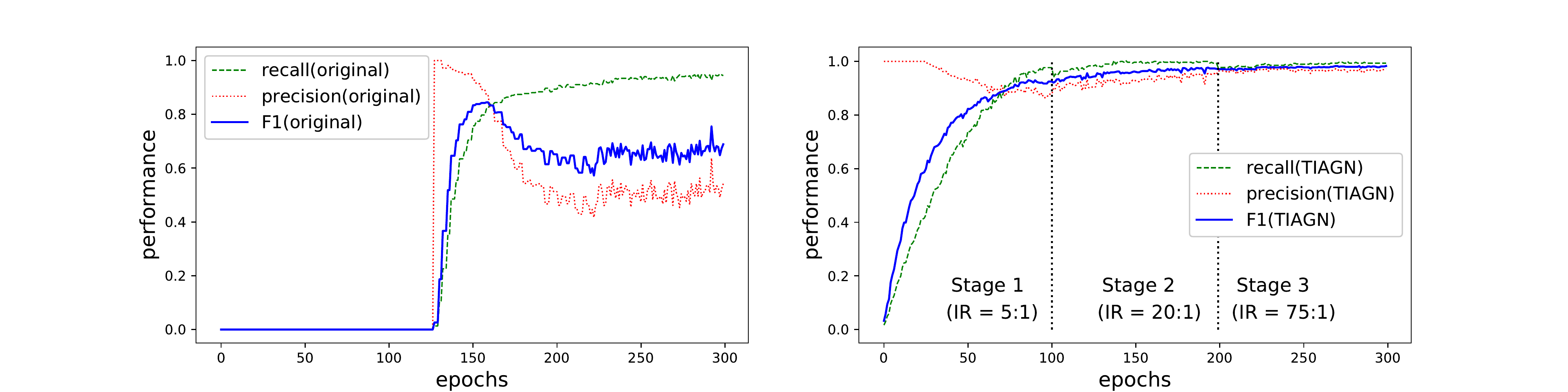}\\
  \caption{The comparing performances between MICNN-1 trained by unaugmented samples (on the left) and MICNN-1 trained by TIAGN samples in the three-stage training process (on the right).}\label{Fig:one_vs_three}
\end{figure}

What the results would be like for MICNN trained only on the unaugmented data with sole stage instead of three stages?
Fig. \ref{Fig:one_vs_three} shows the comparing performance in the training process between the unaugmented dataset (namely, Training 0) and TIAGN samples. The results have been evaluated on the validation set using three metrics: recall, precision, and $F_1$ score. It can be seen that there are at least two visible advantages of the three-stage training strategy. First of all, the model was easy to train in the three-stage process, although the precision rate was a bit low at the beginning.
By contrast, the model based on unaugmented data was hard to train, as we can see from Fig. \ref{Fig:one_vs_three} (on the left) that the recall and the precision kept zero in the early stage of training. It implies that MICNN tends to judge all the candidates as non-pulsars when the training data is high class-imbalanced.
Moreover, the MICNN trained with the three-stage training strategy was stable and convergent, and finally achieved satisfactory performance measures. However, the model based on unaugmented data was more likely to fall into overfitting. The decreasing $F_1$ score curve implies that it fails to predict future observations reliably.

\vspace{0.3 cm}
There are several reasons for the success of MICNN, including the multiple inputs, the augment technique of TIAGN, and the three-stage training strategy. For one thing, multiple inputs to the MICNN play an important role, as the extra features from folded profiles and the DM curves have been shown to help improve the performance of the models (Table \ref{Tab:threestage}). Besides, the TIAGN of data augmentation technique makes our model easier to get its invariant features. Last but not the least, the three-stage training strategy enables MICNN to be trained in a highly class-imbalanced set. These together provide a practical scheme for pulsar candidate sifting and enable the application of the machine learning to a class-imbalanced survey from the next-generation radio telescopes.

\section*{acknowledgements}
\begin{acknowledgements}
      Authors are grateful for supportings from the National Natural Science Foundation of China (grant No: 61075033, 61273248, 11973022), and the Natural Science Foundation of Guangdong Province (No. 2020A1515010710).
\end{acknowledgements}

\bibliographystyle{aasjournal}



\end{document}